\documentstyle[12pt,epsf]{article}
\begin{document}
\setlength{\baselineskip}{0.30in}

\begin{flushright}
UM - TH - 95 - 33\\

December 27, 1995\\
hep-ph/9512433
\end{flushright}

\begin{center}
\vglue .06in
{\Large \bf {Power Corrections in QCD: \
A Matter of Energy Resolution.}}\\[.5in]

{\bf R. Akhoury and V.I. Zakharov}\\
[.05in]

{\it{The Randall Laboratory of Physics\\
University of Michigan\\
Ann Arbor, MI 48109-1120}}\\[.15in]

\end{center}
\begin{abstract}
\begin{quotation}
We consider power-like corrections in QCD which can be viewed as
power surpressed infrared singularities. We argue that the presence
of these singularities depends crucially on the energy resolution.
In case of poor energy resolution, i.e., inclusive cross sections,
there are constraints on infrared singularities expressed by
the Kinoshita-Lee-Nauenberg (KLN) theorem. We rewrite the theorem in
covariant notations and argue that the KLN theorem
implies the extension of the Bloch-Nordsieck cancellation of
logarithmic singularities to the case of linear corrections.

\end{quotation}
\end{abstract}
\newpage


The emission of soft particles in gauge theories exhibits
remarkable regularities. In particular, the spectrum of infrared
photons is universally proportional to $d\omega /\omega$.
Such a spectrum implies a logarithmic divergence
upon integration over the photon energy $\omega$. Moreover, the divergence is
cancelled if the emission cross section is summed up with
radiative corrections to the process without a soft photon; the famous
Bloch-Nordsieck cancellation \cite{bn}.
As a result the physical cross section can
contain only the log of the physical energy resolution, $\Delta E$.

The general nature of this cancellation is revealed by
a quantum-mechanical theorem, due to Kinoshita, Lee and Nauenberg (KLN)
\cite{kln}, according to which all infrared singularities
are canceled provided that summation over all degenerate in energy
states is performed.
An important point is that summation over both initial and final states
is required:
\begin{equation}
\sum_{i,f}|S_{i\rightarrow f} |^2~\sim~free~ of~ singularities
\end{equation}
where $S_{i\rightarrow f}$ are the elements of the $S$-matrix.
Since the summation over initial states
does not correspond to an experimental resolution,
infrared singularities persist, generally speaking,
in physical cross sections. From this point of view
the Bloch-Nordsieck
cancellation, upon the summation over final
states alone, looks as an exception rather than a rule. The reason for this
exception is that in the limit of vanishing photon energy,the
emission and absorption of a photon are indistinguishable.

In QCD, the problem of infrared singularities is compounded due to
the essential presence of collinear singularities , and as such,
 infrared sensitive quantities are affected by
confinement and cannot be calculated reliably. One is therefore
led to consider observables that are infrared safe or those
in which the long distance effects can be isolated into a universal factor
\cite{stcol}.
In the latter approach,the effect of the collinear
singularities is then embedded into phenomenological structure functions
$f_a^h(x)$ where $x$ is the momentum fraction of the parent hadron $h$ carried
by parton $a$. In particular, the cross section for the Drell-Yan (DY)
process $h_1+h_2\rightarrow \mu^+\mu^- + X$ is given by \cite{stcol}:
\begin{equation}
{d\sigma\over dQ^2}(\tau,Q^2)~=~\sum_{a,b}\int_0^1dx_1\int_0^1 dx_2
\int_0^1dx\delta(\tau-x_1x_2x)
f^{h_1}_a(x_1)f^{h_2}_b(x_2)(\sigma_{0}
W_{ab}(x,Q^2)) \label{dyf}
\end{equation}
where,
\begin{equation}
\sigma_{0}= \tau{{4\pi\alpha_{QED}^{2}}\over{9Q^4}}.
\end{equation}
 $S,Q^2$ are squares
of the hadronic and leptonic invariant masses, respectively,
$\tau=Q^2/S$ and $W_{ab}(x,Q^2)$ is the
appropriately normalized (hard) cross section
for $a+b\rightarrow \mu^+\mu^-+partons$, and $x=Q^2/s$ with
$s$ the invariant mass squared
of the partons $a,b$. The structure functions can be deduced from another
process, say, deep inelastic scattering.
As an example of the first approach, one is led to the
construction of infrared safe variables such as thrust $T$ in $e^+e^-$
annihilation:
\begin{equation}
T~=~max_{{\bf n}}{\Sigma ({\bf p}_i{\bf n})\over
\Sigma |{\bf p}_i |}\label{thrust}
.\end{equation}
where ${\bf p}_i$ are the momenta of particles while ${\bf n}$ is a unit
vector.

Most recently, the presence of  linear terms, $\sim \Lambda_{QCD}/Q$
where $Q$ is a large mass parameter, has attracted attention
\cite{cs,wb,az,ks}. These terms do not jeopardize the calculability
of various observables but provide us with a measure of their
infrared sensitivity.
Consider, for example, the emission of a soft gluon in the case of thrust:
\begin{equation}
(1-T)_{soft}~\sim~\int^{\Lambda_{QCD}}_0
{ {d\omega\over \omega} {\omega\over Q} \alpha_s(\Lambda^2_{QCD})}
{}~\sim~{\Lambda_{QCD}\over Q}\label{linear}
\end{equation}
where $\Lambda_{QCD}$ is an infrared cut off such that $\alpha_s
(\Lambda_{QCD}) \sim 1$, $Q$ is the total energy and the factor
$\omega/Q$ is due to the definition of the thrust (\ref{thrust}).
Eq (\ref{linear}) clearly demonstrates the presence of linear terms in
thrust and that they arise due to soft gluons. This is a general feature
\cite{az,ks}.
Remarkably enough, recent developments indicate a universality
of these terms to all orders in
the large coupling $\alpha_s(\Lambda_{QCD})$ \cite{az,ks},
elevating thier status to that of the logarithmic divergencies.
The statement
on the universality of linear corrections is formulated, however,
in less transparent terms.
Two ingredients are essential:the renormalon technique
\cite{mueller} and use of resummed cross sections \cite{cs,ks}.
In particular, in case of thrust one can show \cite{ks} that
\begin{equation}
\langle e^{-\nu (1-T)} \rangle _{1/Q}~=~
e^{-\nu E_{soft}}\label{largenu}
\end{equation}
where $\nu$ is a (large) parameter and the universal quantity $E_{soft}$
is expressed in terms if the cusp anomalous dimension $\gamma_{eik}$:
\begin{equation}
E_{soft}~=~{1\over Q}\int {dk^2_{\perp}\over k^2_{\perp}}
\gamma_{eik}(\alpha_s(k_{\perp}))k_{\perp}
.\end{equation}
The anomalous dimension $\gamma_{eik}$
controls various hard cross sections
and is calculable perturbatively \cite{rad}.
Moreover, a similar $1/Q$ piece can be identified
in the DY cross section \cite{cs,az,ks}.
Thus, linear terms appear to be no less universal than the leading log
corrections,
the technical complications being due to the non-abelian nature of
gluons.

In this note we will present arguments that linear terms
and soft logarithmic divergencies share not only
the property of universality but in some cases, a
Bloch-Nordsieck type of cancellation as well. Namely, if one considers an
inclusive cross section, that is, a case of poor energy resolution, the
linear terms cancel. If, on the other hand, the accuracy of measurements
on the final state is of the order of an infrared parameter, then linear terms
survive. As an example of an inclusive cross section
we will consider the DY process. Observables which assume
precision measurements are exemplified by thrust.
We have explicitly verified our arguments for one- and two-loop abelian
gluons and details will be presented elsewhere \cite{az1}.
We believe that our strategy
and techniques can be generalized to a formal all orders proof.
Our search for a Bloch-Nordsieck type of cancellation was stimulated
in fact by the
results of Ref. \cite{bb} where it was shown that if one uses a photon mass
$\mu$ as an infrared regulator then linear in $\mu$ terms cancel from the
DY cross section at the one-loop level.
We are establishing a general principle behind this apparently
accidental one-loop cancellation. As for observables which
assume precision measurements like thrust, the presence of
$1/Q$ terms in these has been confirmed and the KLN based
arguments presented below are inapplicable.

Thus, we will next consider the $1/Q$ corrections to the DY crossection
for large $\tau$,$(\tau\rightarrow1)$, for which the dominant partonic
process is $q\bar{q}$ annihilation and in which the emitted gluons
are soft with thier energy bounded by $Q(1-\tau)$.
 Our analysis hinges crucially
on the KLN theorem. To be precise, consider the quantity:
\begin{equation}
P_{mn}~=~{1\over m!~ n!}\sum_{i,f}|M_{mn}|^2\label{mn}
\end{equation}
where $M_{mn}$ is the amplitude for the process
$q+\bar{q}+m~gluons\rightarrow\gamma^{*}+n~gluons$.
In general, $P_{mn}$ contains contributions from disconnected diagrams
of $M_{mn}$. We will refer to $P_{mn}$ as the Lee-Nauenberg probabilities.
The assertion of the KLN theorem is that the quantity P,
\begin{equation} P~=~\sum_{m,n}P_{mn}
\end{equation}
contains no infrared sensitivity. In particular, and what is relevant for our
discussion, it does not contain terms linear
in an infrared cut off $\lambda$ (which among other possibilities,can be
$(k_{\perp})_{min}$,or the mass of a U(1) gauge boson).

Keeping in mind that the linear terms originate due to soft radiation,
let us consider the case of single soft gluon emission in the DY process.
We can have, therefore, the diagrams with the soft gluon
in the final state , see Fig. 1(b). Arrows on the gauge boson line
denote the direction of momentum flow
and we  have also indicated the unitarity cut. The open
circles denote the interaction with the $\gamma^{*}$ which is
not explicitly considered.
The set of all such diagrams will be denoted by $P_{01}$.
Moreover, according to the KLN prescription for degenerate states, we must
consider diagrams where the soft gluon is absorbed in the initial state.
Thus  we are led to consider in $P_{mn}$ of eq (\ref{mn}) absorption amplitudes
squared  which correspond to to the digrams in Fig. 1(c).
We denote the set of all such diagrams by $P_{10}$. Note that superficially,
$P_{10}$ and $P_{01}$ look similar, however, the energy momentum constraints
are different for the two sets. Finally, we have diagrams with virtual
gluons alone. In fact, we can include the hard interactions as well as
virtual gluons into a single blob and
 we will follow this notation (hatched region in the figures). Thus the
diagram in Fig. 1(a)  will be denoted by $P_{00}$
and includes the contribution
of the virtual gluons.

In addition to the above, we also have the contribution of the disconnected
diagrams. From the
interference between the connected and
disconnected diagrams , we can get contributions to $P_{mn}$ from
sets of the type in Figs.1(d), 1(e) . These sets will be denoted generically by
$P_{11}^{(1)}$. Notice that the quark and antiquark propagators are the same in
$P_{00}$ as in the corresponding $P_{11}^{(1)}$. In fact it is easy to check
that
$P_{11}^{(1)}$ can be obtained from $P_{00}$
by the  replacement of the propagator of the virtual gluon
by $-2\pi\delta(k^2)$.
The contribution from disconnected diagrams do not end here. In fact
an infinite number of disconnected
gluon lines can be added without changing the
order of perturbation theory.
An analysis shows, and this is one of our main results,
that after a suitable rearrangement of the perturbation series, the
contribution of the disconnected pieces can be
multiplicatively factored out:
\begin{equation}
P~=~\sum_{m,n}P_{mn}~=~\sum_{m,n}\sum_{\alpha}
\left({1\over (m-\alpha )!~(n-\alpha )!}D_{d}(m-\alpha ,n-\alpha) \right)
\cdot \left( P_{00}+P_{01}+P_{10}+ \tilde P_{00} \right).\label{klnsum}
\end{equation}
Here, $D_{d}(m-\alpha ,n-\alpha )$ denotes the disconnected
Green's function with $(m-\alpha )$ ingoing and
$(n-\alpha)$ outgoing gluons. Moreover, $\tilde P_{00}$ is identical to
$P_{00}$ in all respect but one. Namely, the gluon propagator
of $P_{00}$ is replaced by its complex conjugate:
\begin{equation}
-2\pi\delta (k^2)+
{i\over k^2+i\epsilon}~=~{i\over k^2-i\epsilon}~=~
\left({-i\over k^2+i\epsilon}\right)^*.
\end{equation}
Thus, $\tilde P_{00}$ cannot
 be evaluated by applying the standard Feynman rules.

Now, from the KLN theorem we know that the sum
$(P_{00}+P_{01}+P_{10}+\tilde{P_{00}})$
does not contain any IR sensitive terms. Note that the
physical Bloch-Nordsieck crossection corresponds to the
sum $P_{00}+P_{01}$.
Let us first briefly consider the soft logarithmic divergences.
It is easy to see that in the leading order as the gluon
energy vanishes,then for the log divergent terms,
$P_{01} \sim P_{10}$ and for the virtual corrections to this
accuracy we may replace the gluon propagator in
both $P_{00}$ and $\tilde P_{00}$ by the $\delta(k^2)$ piece.
Thus from the KLN theorem we conclude that $P_{00}+P_{01}$ is
free of soft logarithmic divergences, which is just the statement
of the Bloch-Nordsieck cancellation at this accuracy. We
would like to extend this argument to the linear terms
in the DY crossection as well.

Specifically, we identify the soft component of
$\sigma_{0}W_{q\bar{q}}(x)$ in eq.(\ref{dyf})
for large $\tau$
with the inclusive probabilities $P_{00}+P_{01}$ above.
Then we take the moments
with respect to $\tau$
 of eq.(\ref{dyf}), upon which, it reduces to a product of moments
of the structure functions and of $\sigma_{0}W_{q\bar{q}}$. The
moments of this latter quantity are with respect to $x$, which
in turn is related to the energy of the emitted gluon in the
soft region (large moments) by \cite{stcat}
\begin{equation}
\omega=Q(1-x)/2.
\end{equation}
If, for example,we denote the transverse momentum cutoff of the
emitted gluon by $\lambda$ then $x$ is bounded by
$(1-2\lambda/Q)$.
Thus, linear terms in the cut-off do not arise from
the virtual corrections in DY. Then we see from eq(\ref{klnsum})
that the quantity $P_{01}+P_{10}$ does not contain terms linear
in $\lambda$ (upon taking moments, which is henceforth understood).

We will next show that it follows from the Low theorem \cite{low}
that
\begin{equation}
P_{01}-P_{10}~=~O(\lambda ^2)~+~finite~ terms~independent~of~\lambda
.\end{equation}
Then since the virtual gluon diagrams cannot contribute
linear terms in $\lambda$
it follows  from the above that $P_{01})$
has infrared sensitivity of $O(\lambda ^2)$ .

 We can show using the Low theorem that to linear accuracy:
\begin{equation}
P_{01}~=~
(\sum|T|^2){2\alpha_s \over \pi}
\int{d^3k\over 2k_0}\delta \left( (p_1+p_2-k)^2-Q^2\right)
\cdot {2Q^2s\over 2(p_1\cdot k)2(p_2\cdot k)}
\end{equation}
where, $p_1,p_2$ are the momenta of the quark (antiquark) rspectively and $T$
is the radiationless amplitude,i.e., for the process
$q+\bar{q}\rightarrow \gamma^*\rightarrow l^+l^-$. To see this, we note that
the amplitude $M_{01}$
can be expressed upto terms $\omega^0$ as:
\begin{equation}
M_{01}^{\alpha} ~=~g_s{(2p_2 -k)_{\alpha}\over 2p_2\cdot k- k^2 }T
- g_s{(2p_1-k)_{\alpha}\over 2p_1\cdot k-k^2}T
+g_s(p_1+p_2)\cdot k
\left( {p_{1\alpha}\over p_1\cdot k}-{p_{2\alpha}\over p_2\cdot k}\right)
{\partial T\over \partial s}
\end{equation}
where, $s\equiv p_1\cdot p_2 $ and $g_s^{2}/(4\pi)=\alpha_{s}$.
Next, in calculating $\sum |M_{01}|^2$ we keep terms
$O(1/k^2)$ and $O(1/k)$ and in this way arrive at the expression
given above. Similar expressions can be obtained for
$P_{10}$ and we find thier moments to be
 equal for a fixed $Q^2$, to this accuracy.

Thus, we conclude that  the
DY cross section, has no terms linear in the infrared cut off $\lambda$.
The universal term is cancelled from the inclusive cross section
as was observed first by an explicit calculation in Ref \cite{bb}.

The approach of the present paper can be generalized to higher orders.
We have checked explicitly \cite{az1} that at least to two loops
in a theory with abelian gluons all steps indeed generalize.
In particular, the observation (\ref{klnsum}) that the
KLN sum (\ref{mn}) involves in fact both standard and complex conjugated
gluon propagators remains true and evolves into a pattern which
can be readily generalized to any order.
Thus, the expression for the probability $P$ for the case of
two gluons again assumes a form in which the disconnected pieces
may be factorized out, i.e.,
\begin{eqnarray}
P~=~\sum_{m,n}P_{mn}~=~\sum_{m,n}\sum_{\alpha}
\left({1\over (m-\alpha )!~(n-\alpha )!}(\alpha+1)
D_{d}(m-\alpha ,n-\alpha) \right) \nonumber \\
\cdot \left( P_{00}+P_{01}+P_{02}
+P_{10}+P_{20}+ \tilde{\tilde P_{00}}
+P_{11}^{(2)}+ \tilde P_{10}+ \tilde P_{01}+ \tilde P_{00} \right), \label{s2}
\end{eqnarray}
where, to this order, $P_{01}$ contains the square of the amplitude
with  one virtual and one real
emitted gluon, and $\tilde P_{01}$ is obtained from the former
by symmetrically replacing the virtual gluon line  in the amplitude
by its complex
conjugate. Similarly for $P_{10}$ and $\tilde P_{10}$ which apply
for the absorbed gluon. $\tilde{\tilde P_{00}}$
 is obtained from
$P_{00}$ by replacing both the virtual gluon lines in the
 corresponding amplitude
by thier complex conjugates, whereas $\tilde P_{00}$ is obtained
by symmetrically replacing only one. $P_{02}$ refers to the
probability for the emission of two gluons and $P_{20}$ to the
corresponding absorbtion amplitude.$P_{11}^{(2)}$ is the probability
for the emission of one gluon and the absorbtion of another.
Next the Low theorem may be generalized to the case of two abelian
soft gluons and this can be used to relate the
corresponding emission and absorbtion probabilities to linear accuracy.
Proceeding in a manner analogous to the one gluon case, we can
then show again that the KLN theorem implies the Bloch-Nordsieck
cancellation of terms both logarithmic and linear in $\lambda$.
Moreover, at least in the theory with abelian gluons, the factorization
of the linear terms as well, in the
 inclusive probabilities is  manifested \cite{na}.
Details will be given elsewhere \cite{az1}.

To summarize, we have suggested a criteria  for the presence of $1/Q$
corrections in observables. For those having poor energy resolution
 we have presented arguments that combining
the KLN and Low theorems
provides a powerful means of proving a Bloch-Nordsieck type of cancellation
valid to linear approximation in an infrared cut off.
 Let us also mention that
the KLN theorem allows for a compact derivation of the
cancellation of power-like corrections in inclusive weak decays
found explicitly in Ref \cite{bbz} in the one loop order. Moreover,
the generalization to higher loops is straightforward \cite{az1}.

We are grateful to M. Beneke for communicating to us the results of
ref.\cite{bb} before publication and we would like to thank
 T.D. Lee and A.H. Mueller for illuminating discussions.


\newpage

\begin{figure}
\hbox to \hsize{ \hfil \epsffile{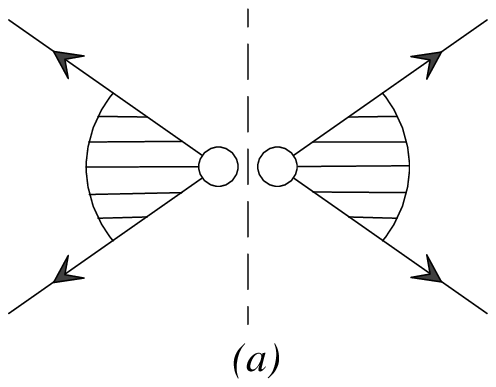} \hfil }
\hbox to \hsize{ \epsffile{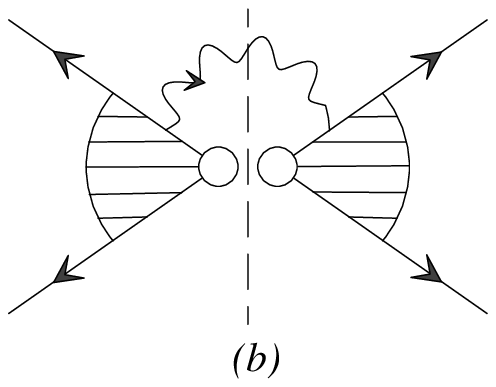}\hfil \epsffile{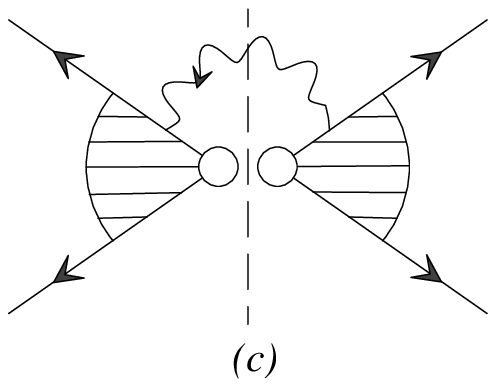}}
\hbox to \hsize{ \epsffile{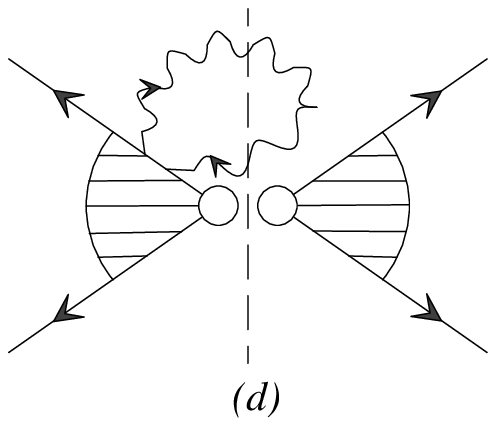}\hfil \epsffile{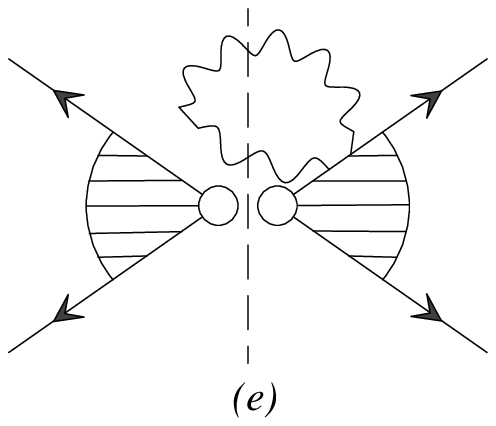}}
\caption{ Cut diagrams representing (a) $P_{00}$,(b)$P_{01}$,(c)
$P_{10}$,and (d) $+$ (e)$P_{11}^{(1)}$.}
\label{Fig.1}
\end{figure}


\begin{thebibliography}{99}
\bibitem{na}The extension of the above arguments to the non-abelian
case though not straightforward is quite plausible. The expressions
for $P$, eq.(\ref{klnsum}), and eq.(\ref{s2})
of the text, are valid here as well. Since for the
case discussed in the letter,the pattern for the linear terms
follows closely that for the logarithmic ones, we expect
that for the inclusive cases under consideration, the rest
of the argument will generalize as well.
\end{thebibliography}
\end{document}